\documentclass[11pt]{article}

\input epsf.sty

\def\lromn#1{\uppercase\expandafter{\romannumeral#1}}

\usepackage{graphicx,amsmath,amssymb}
\def\lromn#1{\uppercase\expandafter{\romannumeral#1}}

\begin{document}
\begin{flushright}
\today \\
\end{flushright}

\vspace{2cm}
\begin{center}
\begin{large}
{\bf Thorium isomer for radiative emission
of neutrino pair}
\end{large}

\vspace{2cm}

 N. Sasao$^{\dagger}$, S. Uetake, A. Yoshimi$^{\dagger}$, K. Yoshimura$^{\dagger}$ and
M. Yoshimura

\vspace{0.5cm}
Center of Quantum Universe, Faculty of
Science, Okayama University \\
Tsushima-naka 3-1-1 Kita-ku Okayama
700-8530 Japan

$^{\dagger}$
Research Core for Extreme Quantum World,
Okayama University \\
Tsushima-naka 3-1-1 Kita-ku Okayama
700-8530 Japan \\

\end{center}

\vspace{5cm}

\begin{center}
\begin{Large}
{\bf ABSTRACT}
\end{Large}
\end{center}

It is proposed to use the isomer ionic ground state $^{229m}$Th$^{4+}$
embedded in transparent crystals for precision determination of unknown neutrino
parameters.
Isolation from solid environment of the proposed nuclear process, along with
available experimental techniques of atomic physics, has a great potentiality
for further study.

\vspace{5cm}

Key words

Neutrino mass, Majorana particle,
Thorium isomer, Macro-coherence

\newpage
{\bf Introduction}
\hspace{0.2cm}
Despite of remarkable success of neutrino oscillation
experiments in recent years \cite{nu oscillation data}
a few critically important neutrino parameters 
have been left untouched.
To improve the situation we proposed \cite{my-prd}, \cite{ptep overview} to use atoms,
instead of conventional nuclear target \cite{tritium}, \cite{nu0 beta},
for the purpose of determination of undetermined important
neutrino parameters such as
the smallest neutrino mass, the mass type
(Dirac vs Majorana), and their CP properties, not easy to
access by other means (due to the energy mismatch
to expected neutrino masses of a fraction of eV).

We propose in the present work yet another possibility;
use of the nuclear isomer of thorium ion, $^{229m}$Th$^{4+}$.
This nuclear isomer has an exceptionally small excitation energy
as nuclear levels,
the best estimated value 7.8 $\pm$ 0.5 eV at present \cite{beck-07},
and a small decay rate of nuclear magnetic dipole transition
to ground state yet to be determined.
Due to protection of many surrounding electrons,
nuclear isomer in thorium atom has an advantage of excellent isolation
from environmental effects.
It therefore has spurred great interest both in application and in
fundamental physics:
a possibility of the next-generation frequency standard
\cite{peik-tamm}, 
\cite{CJCampbell:PRL2009},
\cite{WGRellergert:PRL2010},
\cite{GAKazakov:arXiv2011},
and the precision measurement of variation of
fundamental constants \cite{flambaum}.

The process we use for the neutrino mass spectroscopy is atomic de-excitation:
radiative emission of neutrino pair (RENP) from
the ground ion atom of the isomer nucleus (isomer ground state)
to the ground state ion
of the normal (ground) nucleus, denoted by
$|e\rangle \rightarrow |g\rangle + \gamma +\nu_i\nu_j$
where $\nu_i, i=1,2,3$ is the neutrino mass eigen-state.
Location of pair emission thresholds,
$\omega_{ij} = \epsilon_{eg}/2 - (m_i+m_j)^2/(2\epsilon_{eg})$, 
and the photon spectrum in their vicinity gives
excellent opportunities of determining all neutrino  mass values.
Separation into mass eigen-state thresholds is made possible by precise frequencies
of trigger lasers, an essential ingredient of
RENP process developing field storage along with
target polarization.

Since we need a large number of target atoms, of order the Avogadro
number, for RENP, a  solid environment is the best target choice.
Isolation of the proposed isomer ion from solid environment
is a great advantage in this regard, since we need
a high degree of coherence, normally difficult to
achieve in solids, for developing
a macro-coherent state of target and the stored field
for RENP \cite{yst-pra}, \cite{ptep overview}.

\vspace{0.5cm}
{\bf Th$^{4+}$ ion in transparent solid}
\hspace{0.2cm}
We consider embedding the isomer Th in transparent crystal
such as CaF$_2$.
The initial RENP state $|e \rangle $ is taken as the isomer ion
ground state $^{229m}$Th$^{4+}$,
while the final state $|g\rangle$ is the normal ground state $^{229g}$Th$^{4+}$.
The  embedded $^{229m}$Th$^{4+}$ is a closed shell ion 
of configuration 6p$^6$.
CaF$_2$ solid
has a large electric crystalline field at the ionic site,
reaching field gradient $\sim O(5\times 10^{18}) $V/cm$^{-2}$ \cite{kazakov}.
The large field gradient causes large energy shifts
and large state mixing among parity even or among parity odd states.
The proposed RENP process that occurs within a system of nucleus + core electrons 
is however insensitive to state mixing among different parity states.

\vspace{0.5cm}
{\bf Weak interaction of neutrino pair emission from electron}
\hspace{0.2cm}
According to the standard electroweak theory,
neutrino interaction with atomic electron is described by
 $H_{w} = \langle n | \int dx^3  {\cal H}_{2\nu}^e(x) | n' \rangle$ with
\begin{eqnarray}
&&
 {\cal H}_{2\nu}^e = 
\frac{G_F}{\sqrt{2}} \left( 
\bar{\nu}_e \gamma^{\alpha} ( 1- \gamma_5) \nu_e
\bar{e}  \gamma_{\alpha} ( 1- \gamma_5) e
- \frac{1}{2}
\sum_{i}\bar{\nu}_i \gamma^{\alpha} ( 1- \gamma_5) \nu_i
\bar{e} \gamma_{\alpha} (1- 4 \sin^2 \theta_W - \gamma_5) e
\right)
\,.
\label{2nu vertex}
\end{eqnarray}
Since atomic electrons are described in the non-relativistic regime,
we may ignore terms suppressed by the factor $1/m_e$.
In terms of two component spinors,
\begin{eqnarray}
&&
{\cal H}_{2\nu}^e = 
\frac{G_F}{\sqrt{2}} \left(
e^{\dagger} e \sum_{ij} b_{ij} \nu_j^{\dagger}\nu_i
+ e^{\dagger}\vec{\sigma} e \cdot\sum_{ij}a_{ij} 
 \nu_j^{\dagger}\vec{\sigma}\nu_i
\right) + O(\frac{1}{m_e})
\,, 
\\ &&
b_{ij} = U_{ei}^*U_{ej} - 
\frac{\delta_{ij}}{2} (1 - 4\sin^2 \theta_W )
\,, \hspace{0.5cm}
a_{ij} = - U_{ei}^*U_{ej} + \frac{1}{2}\delta_{ij}
\,,
\label{electron current}
\end{eqnarray}
where $U_{ei}$ is the mixing 
amplitude of $\nu_e$ with the i-th neutrino mass eigen-state.
For neutrino pair emission from core electrons,
the relevant current arises from the 
zero-th component of 4-vector current.
This gives rise to pair emission of the
same species of neutrinos, with $ b_{i} e^{\dagger} e\,, b_i = b_{ii}$,
Our process is insensitive to
Majorana CP phases which appear in off-diagonal elements of
$U_{ei}^*U_{ej}$.
In a recent work \cite{ys-13} this type of mono-pole
contribution, both from inner core electrons and from nucleus,
has been considered and found to greatly enhance RENP rates.

\vspace{0.5cm}
{\bf Coulomb assisted RENP}
\hspace{0.2cm}
We consider RENP in which the photon is emitted from the nuclear isomer and
the neutrino pair from atomic core electrons.
This seemingly separated process is bridged by Coulomb interaction
between atomic core electron and nucleus.
The strength of Coulomb interaction is estimated by Thomas-Fermi model
\cite{atomic physics}, 
resulting in  \cite{ys-13},
\begin{eqnarray}
&&
J_C \equiv  \sum_c \langle c | \frac{Z\alpha}{r} | c\rangle = 
1.6 \times \frac{2^{5/3}}{(3\pi)^{2/3}} Z^{7/3} \alpha^2 m_e
\sim 31 {\rm eV} Z^{7/3}
\,.
\end{eqnarray}
For $^{229}$Th $J_C \sim 1.5 \times 10^8$eV.
This matrix element is insensitive to the quantum numbers
of core electrons $c$.

\begin{figure*}[htbp]
 \begin{center}
 \epsfxsize=0.6\textwidth
 \centerline{\epsfbox{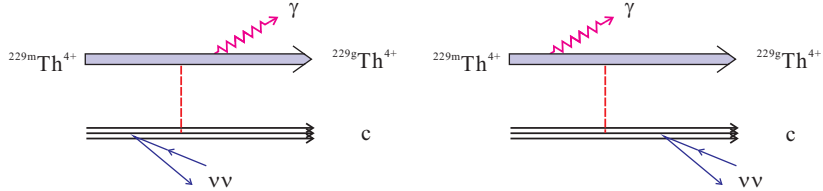}} \hspace*{\fill}
   \caption{Th isomer RENP diagrams.
Electron emitting neutrino pair is one of core electrons.
}
   \label{th renp diagram}
 \end{center} 
\end{figure*}

RENP amplitudes from the isomer ground state
consist of two Feynman diagrams as depicted in
Fig(\ref{th renp diagram}), differing in time sequence of
three interaction vertexes; Coulomb interaction 
sandwiched between nuclear  M1 emission and neutrino pair
emission in the ion ground state.
Two contributions have a common vertex product and differ in energy denominator
in third order perturbation formula.
The vertex product is two neutrino wave functions $\times
J_C g\frac{e}{2m_N}\vec{S}_N\cdot\vec{B}$
where $g$ is the unknown g-factor of isomer M1 transition,
and $\vec{B}$ is the magnetic field of emitted photon.

Energy denominators from these
contribution are given by $2/(\epsilon_m - \omega)^2$ with
$\epsilon_m$ the isomer energy.
Multiplying the remaining factors in amplitude gives Coulomb assisted RENP amplitude,
\begin{eqnarray}
&&
\frac{G_F}{\sqrt{2}}  \sum_i b_{i} \nu_i^{\dagger}  \nu_i
g\frac{e}{2m_N}\vec{S}_N\cdot\vec{B} \frac{2 J_C}{(\epsilon_m-\omega)^2} 
\,.
\end{eqnarray}

\vspace{0.5cm}
{\bf RENP spectral rate}
\hspace{0.2cm}
RENP photon spectral rate is obtained from the squared amplitude,
by summing over neutrino helicities and momenta and
by replacing the emitted photon field strength $|E|^2$ by the extractable
stored field energy density, equal to $\epsilon_{m} n  \times$
a dynamical factor $\eta_{\omega}$.
The quantity $\epsilon_{m} n$ is the energy density  stored 
in the isomer level.
The dynamical factor $\eta_{\omega}(t)$ is time dependent and computed numerically
by solving the master equation for fields and target polarization,
as given in \cite{ptep overview}.
Usually, $\eta_{\omega}$ is much less than unity.
See  below on more of the corresponding physical process.

The neutrino helicity summation has 
the mono-pole current squared term $\frac{1}{2} (1 + \frac{\vec{p}_1\cdot\vec{p}_2}{E_1E_2}
+ \delta_M\frac{m_1 m_2}{E_1E_2} ) $
multiplied by $|b_{i}|^2$. 
Here $\delta_M = 1$ for the Majorana case and zero for the Dirac case.
The phase space integral of neutrino momenta, using the energy-momentum conservation
of 3-body decay, reduces to 
\begin{eqnarray}
&&
\int \frac{d^3 p_1 d^3 p_2}{(2\pi)^2} \delta(E_1 + E_2 + \omega - \epsilon_{eg}) 
\delta(\vec{p}_1 + \vec{p}_2 + \vec{k}) 
\frac{1}{2} (1 + \frac{\vec{p}_1\cdot\vec{p}_2}{E_1E_2}
+ \delta_M\frac{m_1 m_2}{E_1E_2} )
\,, 
\\ &&
= \frac{1}{2\pi \omega}
\int_{E_-}^{E_+} d E_1E_1 E_2
\frac{1}{2} (1 + \frac{\vec{p}_1\cdot\vec{p}_2}{E_1E_2}
+ \delta_M\frac{m_1 m_2}{E_1E_2} )
\,, \hspace{0.5cm}
E_2 = \epsilon_{eg} - \omega - E_1
\,,
\end{eqnarray}
with $E_{\pm}$ derived by 3-body kinematics.
Amplitude squared as described above give quadratic functions of
neutrino energy $E_1$ and the energy integral over $E_1$ can be performed
explicitly.
Explicitly, the spectral shape is given by $4/(\epsilon_m - \omega)^2$ times
\begin{eqnarray}
&&
I(\omega) = 
\sum_{i} |b_i|^2 \Delta_{i}(\omega)
I_{i}(\omega) \theta(\omega_{ii}-\omega)
\,,
\label{rnpe spectrum rate}
\\ &&
I_{i}(\omega) =  
\frac{\omega^2}{3} + \frac{2m_i^2 \omega^2}{3 \epsilon_{eg}(\epsilon_{eg}-2\omega)} 
+ m_i^2 (1 + \delta_M)
\,, \hspace{0.5cm} 
\Delta_{i}(\omega) 
= \left( 1
 -  \frac{4 m_i^2}{\epsilon_{eg} (\epsilon_{eg} -2\omega) }
\right)^{1/2}
\,.
\label{rnpe spectrum rate 3}
\end{eqnarray}

\begin{figure*}[htbp]
 \begin{center}
 \epsfxsize=0.6\textwidth
 \centerline{\epsfbox{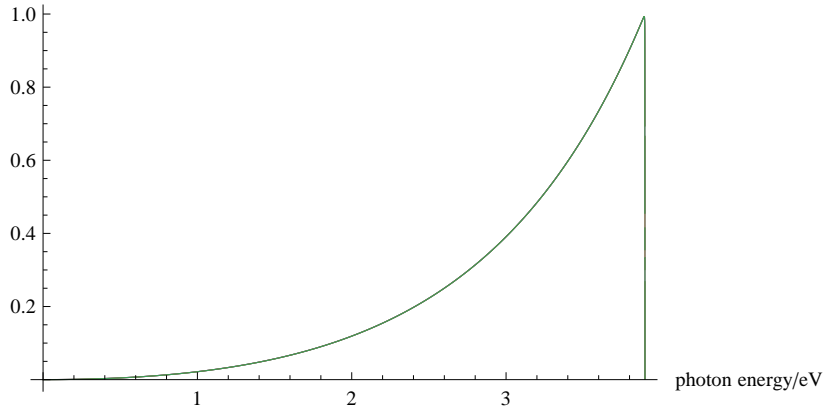}} \hspace*{\fill}
   \caption{Th isomer RENP spectral shape, $4I(\omega)/(\epsilon_m - \omega)^4$ 
in eq(\ref{renp rate used}).
The absolute rate is given by multiplying eq.(\ref{overall rate}).
   Measured neutrino oscillation data are taken into account assuming
   for the smallest neutrino mass 10 meV in the normal (with larger rates)
 and inverted mass hierarchical patterns.
   The Dirac and Majorana case are plotted in different colors,
   but they are indistinguishable with this resolution.
   }
   \label{th spectral shape}
 \end{center} 
\end{figure*}

\begin{figure*}[htbp]
 \begin{center}
 \epsfxsize=0.6\textwidth
 \centerline{\epsfbox{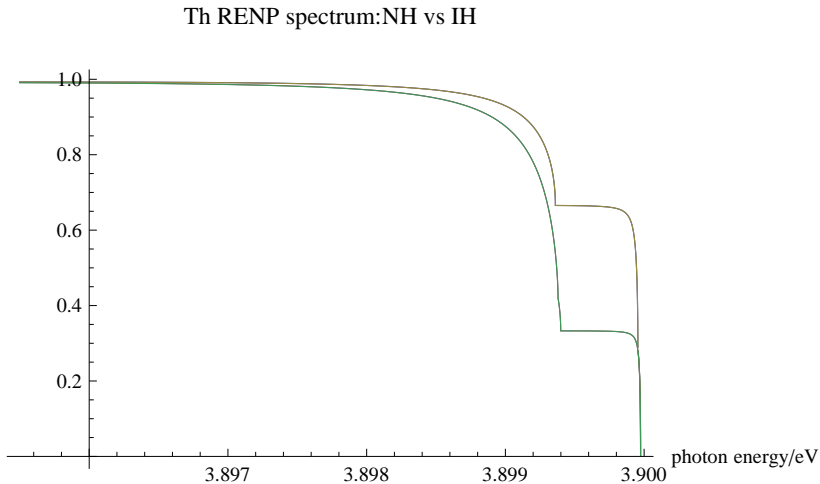}} \hspace*{\fill}
   \caption{Th isomer RENP spectral shape in threshold regions
corresponding to Fig(\ref{th spectral shape}).
NH case in brown and IH case in green.
      }
   \label{th spectral shape 2}
 \end{center} 
\end{figure*}

In Fig(\ref{th spectral shape}) and Fig(\ref{th spectral shape 2}) we illustrate
the spectral shape $I(\omega)$ by taking the isomer energy of
$\epsilon_m = 7.8 $eV.
The spectral rate is decomposed into three factors;
the overall rate, the squared atomic matrix element, and
kinematical factor $I(\omega) $, thus
\begin{eqnarray}
&&
\Gamma_{2\nu \gamma} = \Gamma_n (\frac{2J_C}{(\epsilon_m - \omega)^2})^2
I(\omega) \eta_{\omega}(t)
\label{renp rate used}
\,,
\\ &&
\Gamma_n = \frac{G_F^2}{2} (\frac{g e}{2m_N})^2 n^3 V \epsilon_{eg}
\sim 1.2 \times 10^{-13}{\rm Hz} (\frac{n}{10^{22}{\rm cm}^{-3}})^3 
\frac{V}{{\rm cm}^3} \frac{\epsilon_{eg}}{10{\rm eV}}
\,.
\end{eqnarray}

The actual rate for the thorium isomer, taking 7.8 eV energy,
is $4I(\omega)/(\epsilon_m - \omega)^4$ multiplied by 
\begin{eqnarray}
&&
22 g^2
\,{\rm Hz}  (\frac{n}{10^{21}{\rm cm}^{-3}})^3 \frac{V}{{\rm cm}^3}
\eta_{\omega}(t)
\,.
\label{overall rate}
\end{eqnarray}
The Majorana vs Dirac distinction is difficult for this large
energy spacing $\sim 7.8$eV.
Distinction is improved for smaller energy spacings
of atoms, as discussed in \cite{dpsty-plb}.

\vspace{0.5cm}
{\bf Macro-coherent field storage}
\hspace{0.2cm}
RENP rates depend on the field storage factor $\eta_{\omega}(t)$,
controlled  by two-photon paired super-radiance (PSR) process
$|r\rangle \rightarrow |g\rangle + \gamma \gamma$.
PSR is  triggered by laser irradiation of two frequencies, the one
at $\omega$ and another at $\epsilon_m - \omega > \omega$.
The dynamical factor $\eta_{\omega}(t)$ is defined by a space integrated
quantity of $|E(t, x) (r_1(t,x) - ir_2(t,x)\,)|^2$, expressed
in dimensionless units, and
is calculated by numerically solving the master equation 
for the developed field $E(t,x)$ and target polarization 
$r_1 - ir_2$ of \cite{yst-pra}, \cite{ptep overview}.
It also depends on experimental conditions.

Macro-coherent two-photon PSR is controlled by diagrams
consisting of two vertexes of the isomer M1 and atomic M1.
Atomic M1 vertex is induced by the state mixing of $6p$ electron
in the ion closed shell with electron in $7p$ orbit
caused  by crystalline field. 
Its strength is large due to large $J_C$, 
even if nuclear M1 amplitude $\propto 1/m_N$ is
small.
Thus, both RENP and PSR are assisted  by
Coulomb interaction with nucleus.
Detailed numerical simulation is needed \cite{eta in ptep overview}.

\vspace{0.5cm}
In summary, we proposed to use the nuclear isomer $^{229m}$Th$^{4+}$
embedded in transparent crystals
for the neutrino mass spectroscopy.
It has a merit of excellent isolation from solid environment. 
Thus, although RENP rates are not very large,
a large relaxation time may compensate these moderate rates.

Work on parity violation (PV) effects is in progress.
PV asymmetry along with PV rate is expected to 
be large for $^{229m}$Th RENP.

\vspace{0.3cm}
{\bf Acknowledgements}
\hspace{0.2cm}
This research was partially supported by Grant-in-Aid for Scientific
Research on Innovative Areas "Extreme quantum world opened up by atoms"
(21104002)
from the Ministry of Education, Culture, Sports, Science, and Technology.


\begin{thebibliography}{99}
\bibitem{nu oscillation data}
G. L. Fogli, E. Lisi, A. Marrone, D. Montanino, A. Palazzo, and A. M. Rotunno,
{\it Phys. Rev.} {\bf D 86}, 013012 (2012).

M. C. Gonzalez-Garcia, Michele Maltoni, Jordi Salvado, Thomas Schwetz,
{\it Journal of High Energy Physics} {\bf December 2012}, 123.

D. V. Forero, M. Toacutertola, and J. W. F. Valle,
{\it Phys. Rev.}{\bf D 86}, 073012 (2012).




\bibitem{my-prd}
M. Yoshimura, {\it Phys. Rev.}{\bf D75}.
113007 (2007).

\bibitem{ptep overview}
A. Fukumi et al.,
{\it Progr. Theor. Exp. Phys.}{\bf 2012, 04D002};
arXiv1211.4904v1[hep-ph](2012), and 
and references cited therein.


\bibitem{tritium}
G. Drexlin, V. Hannen, S. Mertens, and C. Weinheimer, {\it Current Direct
Neutrino Mass Experiments},

Advances in High Energy Physics Volume 2013 (2013)Article ID 293986.


\bibitem{nu0 beta}
A. Gando et al,
{\it Phys. Rev. Lett.}{\bf 110}, 062502 (2013), and
arXiv:1201.4664v2[hep-ex] (2012).

M.Auger et al,
 {\it Phys. Rev. Lett.}{\bf 109}, 032505 (2012).


\bibitem{beck-07}
B.R. Beck et al,
{\it Phys. Rev.Lett.}{\bf 98}, 142501(2007).

B.R. Beck, C.Y. Wu, P. Beiersdorfer, G.V. Brown, J.A. Becker, 
K.J. Moody, J.B. Wilhelmy, F.S. Porter, C.A. Kilbourne, 
R.L. Kelley, Proceedings of the 12th International Conference on Nuclear Reaction Mechanisms (2009), 
Varenna, Italy, LLNL-PROC-415170.

\bibitem{peik-tamm}
E. Peik and Ch Tamm,
{\it Europhys. Lett.}{\bf 61},181(2003).

\bibitem{CJCampbell:PRL2009}
C.~J. Campbell, A.~V. Steele, L.~R. Churchill, M.~V. DePalatis, D.~E. Naylor,
  D.~N. Matsukevich, A. Kuzmich, and M.~S. Chapman, Phys. Rev. Lett. {\bf 102},
   233004  (2009).

\bibitem{WGRellergert:PRL2010}
W.~G. Rellergert, D. DeMille, R.~R. Greco, M.~P. Hehlen, J.~R. Torgerson, and
  E.~R. Hudson, Phys. Rev. Lett. {\bf 104},  200802  (2010).

\bibitem{GAKazakov:arXiv2011}
G.~A. Kazakov, M. Schreitl, G. Winkler, J.~H. Sterba, G. Steinhauser, and T.
  Schumm, arXiv:1110.0741  (2012).


\bibitem{flambaum}
V. Flambaum,
{\it Phys. Rev. Lett.}{\bf 97}, 092502(2006).

\bibitem{yst-pra}
M. Yoshimura, N. Sasao, and M. Tanaka,
{\it Phys. Rev}
{\bf A86},013812(2012),
and
{\it Dynamics of paired superradiance},
arXiv:1203.5394[quan-ph] (2012).

\bibitem{kazakov}
G.A. Kazakov, et al,
{\it New Journal of Physics}{\bf 14}, 083019(2012).


\bibitem{atomic physics}
B.H. Bransden and C.J. Joachain, 
{\it Physics of Atoms and Molecules},
2nd edition, Prentice Hall(2003).

\bibitem{ys-13}
M. Yoshimura and N. Sasao,
''{\it Radiative emission of neutrino pair from
nucleus and inner core electrons in heavy atoms}'',
arXiv:1310.6472v1[hep-ph](2013).

\bibitem{dpsty-plb}
D.N. Dinh, S. Petcov, N. Sasao, M. Tanaka,
and M. Yoshimura,
{\it  Phys. Lett.}{\bf B719},154(2012), and  
arXiv1209.4808v1[hep-ph].

\bibitem{eta in ptep overview}
In \cite{ptep overview} a result for
numerical simulation of $\eta_{\omega}(t)$
is presented for pH$_2$ molecule target
(strong source of paired super-radiance (PSR)
of E1 $\times$ E1 transition, and for time dependence see
Fig 14 of this reference).
Its time dependence is complicated:
a fast rise in $O(2$ ns), then a plateau region
of magnitude $O(10^{-2})$ of duration of
several nano-seconds, finally gradual decrease
ending around $10^{-6}$ up to 12 ns (end time of calculation).

Numerical simulations based on the master
equation given in \cite{ptep overview} should be
performed for specific targets considered,
especially for target atoms of weaker PSR sources
and large relaxation time,
which is expected to
give different temporal behaviors and large values. 


\end{thebibliography}
\end{document}